\documentclass[11pt]{article}

\usepackage[latin1]{inputenc}
\usepackage[T1]{fontenc}
\usepackage[english]{babel}
\usepackage{amsfonts}
\usepackage{amssymb}
\usepackage{amsmath}
\usepackage{amsthm}
\usepackage{graphicx}

\newtheorem{proposition}{Proposition}

\newtheorem{teorema}{Theorem}

\newtheorem{remark}{Remark}

\def\be{\begin{equation}}
\def\ee{\end{equation}}
\def\bea{\begin{eqnarray}}
\def\eea{\end{eqnarray}}

\title{The replica symmetric behavior of the analogical neural network}
\author{Adriano Barra\footnote{Dipartimento di Fisica, Sapienza Universit\`a di Roma, Piazzale Aldo Moro 2, 00185, Roma,
Italy},  \ Giuseppe Genovese\footnote{Dipartimento di Matematica, Sapienza Universit\`a di Roma, Piazzale Aldo Moro 2, 00185, Roma,
Italy}, \ Francesco Guerra \footnote{Dipartimento di Fisica, Sapienza Universit\`a di Roma, and INFN, Sezione di Roma 1, Piazzale Aldo Moro 2, 00185, Roma,
Italy}}
\date{November 2009}

\begin{document}
\maketitle

\begin{abstract}
In this paper we continue our investigation of the analogical
neural network, paying interest to its replica symmetric behavior
in the absence of external fields of any type. Bridging the neural network
to a bipartite spin-glass, we introduce and apply a new
interpolation scheme to its free energy that naturally extends the
interpolation via cavity fields or stochastic perturbations to
these models.
\newline
As a result we obtain the free energy of the system as a sum rule,
which, at least at the replica symmetric level, can be solved
exactly. As a next step we study its related self-consistent
equations for the order parameters and their rescaled
fluctuations, found to diverge on the same critical line of the
standard Amit-Gutfreund-Sompolinsky theory.
\end{abstract}

\section{Introduction}\label{due}

The number of  disordered models, whose description is reached
in the frame of  statistical mechanics for complex system, increases year by year
\cite{barabasi}.  As a consequence, the need of powerful tools for
their analysis raises, which ultimately push further the global
field of research suggesting new possible models where their
applicability  can be achieved.
\newline
Among these, interestingly, neural networks have never been
analyzed from an interpolating, stochastic perturbation,
perspective \cite{guerrasg}. In fact, from the early work by
Hopfield \cite{hopfield} and the, nowadays historical, theory of
Amit Gutfreund and Sompolinsky (AGS) \cite{amit,ags1,ags2} to the
modern theory for learning \cite{hotel,peter}, about the
neural networks (thought of as spin glasses with a Hebb-like
``synaptic matrix'' \cite{hebb}) very little is rigorously known.
\newline
Surely several contributions
 appeared  (e.g. \cite{tirozzi1,bovier1,bovier2,bovier3,bovier4,tirozzi2,tirozzi3,talahopfield1,talahopfield2}), often
   following
 understanding of spin-glasses  (e.g. \cite{broken,guerra2,Gsum,pastur,talabook}) and the analysis at low level of stored memories has been achieved.
\newline
However in the high level of stored memories, fundamental
enquiries are still rather obscure. Furthermore general problems
as the
 existence of a well defined thermodynamic limit, achieved for the spin glass case in
 \cite{limterm,limterm2}, are unsolved.
\newline
Previously we introduced an ``analogical version'' of the standard
Hopfield model, by taking the freedom of allowing the learned
patterns to live on the real axes, their probability distribution
being a standard Gaussian $\mathcal{N}[0,1]$ \cite{bg1}.
\newline
Within this scenario, we proved the existence of an ergodic phase
where the explicit expression for all the thermodynamical
quantities (free energy, entropy, internal energy) have been found
to self-average around their annealed expression in the
thermodynamic limit, in complete agreement with AGS theory.
\newline
In this paper, again by using an analogy among neural networks and
bipartite spin glasses, we move on introducing a novel
interpolating technique (essentially based on two different
stochastic perturbations) which we use to give a complete
description of the analogical Hopfield model phase diagram in the
replica symmetric approximation and with high level of stored
memories (i.e. patterns).
\newline
Furthermore we control the fluctuations and correlations of the
order parameters of the theory, whose divergences confirm the
transition line predicted by standard AGS theory to hold even in
this continuous counterpart.
\newline
As a last remark we stress that the whole is exploited without
external fields checking system responses and, as a consequence,
nor retrieval neither the presence of any ``magnetization'' are
discussed and are left for future speculation.
\newline
The paper is organized as follows: In Sec. \ref{uno} we
introduce the analogical neural network with all its statistical
mechanics package of definitions and properties. In Sec.
\ref{due} we analyze its replica symmetric behavior by means of
our interpolating scheme, while in Sec. \ref{tre} we exploit the
fluctuation control to check for regularities and singularities of
the order parameters, obtaining the critical line for the phase
transition from the ergodic regime to a non-ergodic one.
\newline
Sec. \ref{quattro} is left for
 conclusion and outlook.

\section{Analogical neural network}\label{uno}

We introduce a large network of $N$ two-state neurons $\sigma_i=\pm1
$, $i \in (1,..,N)$, which are thought of as quiescent
(sleeping) when their value is $-1$ or spiking (emitting a current
signal to other neurons) when their value is $+1$. They interact
throughout a symmetric synaptic matrix $J_{ij}$ defined accordingly the Hebb
rule for learning,
\be J_{ij} = \sum_{\mu=1}^k \xi_i^{\mu}\xi_j^{\mu}. \ee
Each random variable $\xi^{\mu}=\{\xi_1^{\mu},..,\xi_N^{\mu}\}$
represents a learned pattern and tries to bring the overall
current in the network (or in some part) stable with respect to
itself (when this happens, we say we have a retrieval state, see
e.g. \cite{amit}). The analysis of the network assumes that the
system has already stored $p$ patterns (no learning is
investigated) and we are interested in the case in which this
number increases proportionally (linearly) to the system size
(high storage level).
\newline
In standard literature these patters are usually taken at random
with distribution $P(\xi_i^{\mu})=
(1/2)\delta_{\xi_i^{\mu},+1}+(1/2)\delta_{\xi_i^{\mu},-1}$, while
we extend their support to be on the real axes weighted by a
Gaussian probability distribution, i.e. \be P(\xi_i^{\mu}) =
\frac{1}{\sqrt{2\pi}}e^{-(\xi_i^{\mu})^2/2}. \ee
\newline
The Hamiltonian of the model is defined as follows \be
H_N(\sigma;\xi) = -\frac{1}{N}\sum_{\mu=1}^k
\sum_{i<j}^N\xi_i^{\mu}\xi_j^{\mu}\sigma_i\sigma_j, \ee
which, splitting the summations $\sum_{i<j}^N =
\frac{1}{2}\sum_{ij}^N - \frac12 \sum_i^N \delta_{ij}$ enable us
to write down the following partition function
\begin{eqnarray}\label{zeta} Z_N(\beta;\xi) &=& \sum_{\sigma}
\exp{\Big(\frac{\beta}{2N}\sum_{\mu=1}^k\sum_{ij}^N
\xi_i^{\mu}\xi_j^{\mu}\sigma_i\sigma_j -
\frac{\beta}{2N}\sum_{\mu=1}^k\sum_{i}^N (\xi_i^{{\mu}})^2 \Big)}
\\ \nonumber &=&
\tilde{Z}(\beta;\xi) \times \Big(
e^{\frac{-\beta}{2N}\sum_{\mu=1}^k \sum_{i=1}^N
(\xi_i^{{\mu}})^2}\Big).
\end{eqnarray}
$\beta$, the inverse temperature in spin glass theory, denotes the
level of noise in the network and we defined
\be\label{Ztilde} \tilde{Z}(\beta;\xi)=
\sum_{\sigma}\exp(\frac{\beta}{2N}\sum_{\mu=1}^k\sum_{ij}^N
\xi_i^{\mu}\xi_j^{\mu}\sigma_i\sigma_j ). \ee Notice that the last
term at the r.h.s. of eq. (\ref{zeta}) does not depend on the
particular state of the network.
\newline
As a consequence, the control of the last term easily follows: \be
\ln Z_{N,k}(\beta;\xi)= \ln \tilde{Z}_{N,k}(\beta;\xi) -
\frac{\beta}{2N}\sum_{\mu}^k\sum_{i}^N (\xi_i^{\mu})^2 =  \ln
\tilde{Z}_{N,k}(\beta;\xi) - \frac{\beta}{2}\hat{f}_N \ee where,
as $\hat{f}_N$ is a sum of independent random variables,
$\mathbb{E}\hat{f}_N=k$ and
$\lim_{N\rightarrow\infty}(1/N)\mathbb{E}\hat{f}_N=k/N$, which in
the thermodynamic limit, simply adds a term $-\alpha\beta/2$ to
the free energy (to be defined in (\ref{freeenergy})).
\newline
Consequently we focus just on $\tilde{Z}(\beta;\xi)$. Let us apply
the Gaussian integration \cite{ellis} to linearize with respect to
the bilinear quenched memories carried by the
$\xi_i^{\mu}\xi_j^{\mu}$: The expression for the partition
function (\ref{Ztilde}) becomes (renaming $\tilde{Z} \to Z$ for
simplicity) \be\label{bipartito} Z_N(\beta;\xi) =
\sum_{\sigma}\int \prod_{\mu=1}^k d\mu(z_{\mu})
\exp\Big(\sqrt{\frac{\beta}{N}}\sum_{\mu=1}^k \sum_{i=1}^N
\xi_i^{\mu}\sigma_i z_{\mu}\Big), \ee with $d\mu(z_{\mu})$
standard  Gaussian measure for all the $z_{\mu}$.
\newline
Taken $O$ as a generic function of the neurons, we define the
Boltzmann state $\omega_{\beta}(O)$ at a given level of noise
$\beta$ as \be \omega_{\beta}(O) = \omega(O)=
(Z_N(\beta;\xi))^{-1}\sum_{\sigma}O(\sigma)e^{-\beta
H_N(\sigma;\xi)}, \ee and often we drop the subscript $\beta$ for
the sake of simplicity. The $s$-replicated Boltzmann measure is
defined as $\Omega = \omega^1\times \omega^2 \times ... \times
\omega^s$ in which all the single Boltzmann states are independent
states at the same noise level $\beta^{-1}$ and share an identical
distribution of quenched memories $\xi$. For the sake of
clearness, given a function $F$ of the neurons of the $s$ replicas
and the freedom of using the symbol $a \in [1,..,s]$ to label
replicas, such an average can be written as
\be \Omega(F(\sigma^1,...,\sigma^s)) =
\frac{1}{Z_N^s}\sum_{\sigma^1}\sum_{\sigma^2}...\sum_{\sigma^s}
F(\sigma^1,...,\sigma^s)\exp(-\beta \sum_{a=1}^s
H_N(\sigma^{a},\xi)). \ee
The average over the quenched memories will be denoted by
$\mathbb{E}$ and for a generic function of these memories $F(\xi)$
 can be written as \be \mathbb{E}[F(\xi)] = \int
\prod_{\mu=1}^p \prod_{i=1}^N \frac{d
\xi_i^{\mu}e^{-\frac{(\xi_i^{\mu})^2}{2}}}{\sqrt{2\pi}}F(\xi)=
\int F(\xi)d\mu(\xi), \ee of course $ \mathbb{E}[\xi_i^{\mu}]=0$
and $ \mathbb{E}[(\xi_i^{\mu})^2]=1$.
\newline
We use the symbol $\langle . \rangle$ to mean $\langle . \rangle =
\mathbb{E}\Omega(.)$.
\newline
In the thermodynamic limit, it is assumed
$$
\lim_{N \rightarrow \infty} \frac{p}{N}= \alpha,
$$
$\alpha$ being a given real number, parameter of the theory.
\newline
For the sake of simplicity we allow a little abuse in the notation
so to use the symbol $\alpha$ even at finite $N$, still meaning
the ration among the two parties.
\newline
The main quantity of interest is the quenched intensive pressure
defined as \be\label{freeenergy} A_N(\alpha,\beta)= -\beta
f_N(\alpha,\beta) = \frac{1}{N}\mathbb{E}\ln Z_N(\beta;\xi). \ee
Here, $f_N(\alpha,\beta)= u_N(\alpha,\beta)-\beta^{-1}s_N(\alpha,\beta)$
is the free energy density, $u_N(\alpha,\beta)$ the internal
energy density and $s_N(\alpha,\beta)$ the intensive entropy.

Reflecting the bipartite nature of the Hopfield model expressed by
 eq. (\ref{bipartito})  we introduce two other order parameters:
 the first is the overlap between the replicated neurons (first party overlap), defined
as \be q_{ab}= \frac1N \sum_{i=1}^N \sigma_i^a \sigma_i^b \in
[-1,+1], \ee and the second the overlap between the replicated
Gaussian  variables  $z$ (second party overlap), defined as \be
p_{ab} = \frac1p \sum_{\mu=1}^k z_a^{\mu}z_b^{\mu} \in (-\infty,
+\infty). \ee
Both the two order parameters above play a considerable role in
the theory as they can express thermodynamical quantities
\cite{bg1}.

\section{Replica symmetric free energy}\label{due}

In this section we pay attention to the structure of the free
energy: we want to obtain the latter via a sum rule in which we
may isolate explicitly the order parameter fluctuations so to be
able to neglect them achieving a replica-symmetric behavior.
\newline
Due to the equivalence among neural network and bipartite
spin-glasses, we generalize the way cavity field and the
stochastic stability techniques work on spin glasses to these
structures by introducing a new interpolation scheme as follows:
\newline
For the sake of clearness, in order to exploit the interpolation method adapted to 
the physics of the model, we introduce $3$ free parameters in the
interpolating structure (i.e. $a,b,c$) that we fix a fortiori,
once the sum rule is almost achieved.
\newline
In a pure stochastic stability fashion \cite{guerra2}, we need to
introduce also two classes of i.i.d. $\mathcal{N}[0,1]$ variables,
namely $N$ variables $\eta_i$ and $K$ variables
$\tilde{\eta}_{\mu}$, whose average is still encoded into the
$\mathbb{E}$ operator and by which we define the following
interpolating quenched pressure $\tilde{A}_{N,k}(\beta,t)$
\begin{eqnarray}\label{interpolante} &&\tilde{A}_{N,k}(\beta,t) =
\frac1N\mathbb{E}\log\sum_{\sigma}\int \prod_{\mu}^k
d\mu(z_{\mu})\exp(\sqrt{t}\frac{\beta}{N}\sum_{i,\mu}^{N,k}\xi_i^{\mu}\sigma_i
z_{\mu}) \\ &\cdot& \nonumber \exp(a\sqrt{1-t} \sum_i^N \eta_i
\sigma_i)\exp(b\sqrt{1-t}\sum_{\mu}^k
\tilde{\eta}_{\mu}z_{\mu})\exp(c\frac{(1-t)}{2}\sum_{\mu}^kz_{\mu}^2).
\end{eqnarray}
We stress that $t\in[0,1]$ interpolates between $t=0$ where the
interpolating quenched pressure becomes made of by non-interacting
systems (a series of one-body problem) whose integration is
straightforward and the opposite limit, $t=1$, that recovers the
correct quenched free energy (\ref{freeenergy}).
\newline
The plan is then to evaluate the $t$-streaming of such a quantity
and than obtain the correct expression by using the fundamental
theorem of calculus: \be\label{sumrule} A_{N,k}(\beta) =
\tilde{A}_{N,k}(\beta,t=1) = \tilde{A}_{N,k}(\beta,t=0) + \int_0^1
dt' \Big(\partial_t \tilde{A}_{N,k}(\beta,t)\Big)_{t=t'}. \ee When
evaluating the streaming $\partial_t \tilde{A}$ we get the sum of
four terms $(A,B,C,D)$: each comes as a consequence of the
derivation of a corresponding  exponential term appearing into the
expression (\ref{interpolante}).
\newline
Once introduced the averages $\langle \cdot \rangle_t$ that
naturally extend the Boltzmann measure encoded in the
interpolating scheme (and reduce to the proper one whenever
setting $t=1$), we can write them down as
\begin{eqnarray} \nonumber
A &=& \frac1N \sqrt{\frac{\beta}{N}}
\frac{1}{2\sqrt{t}}\sum_{i,\mu}^{N,k}\mathbb{E}\xi_{i,\mu}\omega(\sigma_i
z_{\mu})
 = \frac{\beta}{2N}\mathbb{E}\sum_{\mu}^k
\omega(z_{\mu}^2) - \frac{\alpha\beta}{2}\langle q_{12} p_{12}
\rangle_t,
\\ \nonumber
B &=& \frac{-a}{2N\sqrt{1-t}}\sum_i^N \mathbb{E} \eta_i
\omega(\sigma_i) = -\frac{a^2}{2} \big( 1 - \langle q_{12}
\rangle_t\big),
\\ \nonumber
C &=& \frac{-b}{2N\sqrt{1-t}}\sum_{\mu}^k \mathbb{E}
\tilde{\eta}_{\mu} \omega(z_{\mu}) =
\frac{-b^2}{2N}\sum_{\mu}^k\mathbb{E}\omega(z_{\mu}^2) +
\frac{\alpha b^2}{2}\langle p_{12} \rangle_t,
\\ \nonumber
D &=& \frac{-c}{2N}\sum_{\mu}^k \omega(z_{\mu}^2),
\end{eqnarray}
where in the first three equations we used integration by parts (Wick theorem).
\newline
In the replica symmetric ansatz, the order parameters do not
fluctuate with respect to the quenched average and the only values
(at any given $\beta,\alpha$ point) they gets are $\langle q
\rangle = \bar{q}, \langle p \rangle = \bar{p}$, where the bars
denote the replica symmetric approximation.
\newline
Summing all the contributions $(A,B,C,D)$ and adding and
subtracting the term $\alpha \beta\bar{q}\bar{p}/2$ (that we use
to center and complete the square of the two overlaps), we get
\begin{eqnarray} \frac{d \tilde{A}_{N,k}(\beta,t)}{dt} &=& (\beta
- b^2 -c)\frac{1}{2N}\mathbb{E}\sum_{\mu}^k\omega(z_{\mu}^2)
- \frac{\alpha \beta}{2} \langle q_{12} p _{12} \rangle_t - \\
\nonumber &-& \frac{a^2}{2}(1 - \langle q_{12} \rangle_t) +
\frac{\alpha b^2}{2}\langle p_{12} \rangle_t + \frac{\alpha
\beta}{2}\bar{q}\bar{p} - \frac{\alpha
\beta}{2}\bar{q}\bar{p}.\end{eqnarray} So we see that if we choose
$$
a = \sqrt{\alpha \beta \bar{p}}, \ \ b = \sqrt{\beta \bar{q}} \ \
c = \beta (1 - \bar{q}),
$$
we get \be\label{streamA} \frac{d \tilde{A}_{N,k}(\beta,t)}{dt} =
-\frac{\alpha \beta}{2}\langle  (q_{12} - \bar{q}) (p_{12} -
\bar{p})\rangle_t - \frac{\alpha \beta}{2} \bar{p}(1-\bar{q}). \ee
Once inserted the expression (\ref{streamA}) into
eq.(\ref{sumrule}) the sum rule for the free energy is achieved.
\newline
In order to get the replica symmetric solution
$A_{N,k}^{RS}(\beta)$ we impose the self-averaging of the
overlaps, so that we need to evaluate only \be
A_{N,k}^{RS}(\beta) = \tilde{A}_{N,k}(\beta,t=0) -\frac{\alpha
\beta}{2} \bar{p}(1 - \bar{q}) -\frac{\alpha \beta}{2},\ee where
the last term at the r.h.s. comes from the diagonal term of the
first party as explained in Sec. \ref{uno}.
\newline
The evaluation of  $\tilde{A}_{N,k}(\beta,t=0)$ is easily
available because it is a one-body calculation, which implies
 factorization in the volume sizes. Namely, we
have to evaluate explicitly the quantity
\begin{eqnarray} &&\tilde{A}_{N,k}(\beta,t=0) = \\ \nonumber && = \frac1N \mathbb{E} \log
\sum_{\sigma} \int \prod_{\mu}^k d
\mu(z_{\mu})e^{\sqrt{\alpha\beta\bar{p}}\sum_i^N\eta_i\sigma_i}e^{\sqrt{\beta
\bar{q}}\sum_{\mu}^k\tilde{\eta}_{\mu}z_{\mu}}e^{\frac{\beta}{2}(1-\bar{q})\sum_{\mu}^p
z_{\mu}^2}  \\ && = \frac1N \mathbb{E} \log \sum_{\sigma}
e^{\sqrt{\alpha \beta \bar{p}}\sum_i^N \eta_i \sigma_i} +
\nonumber \\ && \ \ + \frac1N \mathbb{E}\log\int \prod_{\mu}^k
dz_{\mu} e^{-\frac12\sum_{\mu}^k z_{\mu}^2 (1 -
\beta(1-\bar{q}))}e^{\sqrt{ \beta \bar{q}}\sum_{\mu}^k \eta_{\mu}
z_{\mu}} \nonumber \\
&& = \log 2 + \int d\mu(\eta) \log \cosh\Big( \sqrt{\alpha \beta
\bar{p}} \eta\Big) + \nonumber \\ && \ \ +  \frac{\alpha}{2}\log
\Big( 1 - \beta(1-\bar{q}) \Big) + \alpha \mathbb{E}\log \int dr
e^{-r^2/2} e^{\sqrt{\frac{\beta \bar{q}}{1-\beta(1-\bar{q})}}\eta
r}, \nonumber
\end{eqnarray}
where we introduced $r= \sigma z$, $\sigma$ defining  the standard
Gaussian variance such that \be \sigma^2 =
(1-\beta(1-\bar{q}))^{-1}.\ee
\newline
As a consequence we get
\begin{eqnarray} \tilde{A}_{N,k}(\beta,t=0) &=& \log 2 + \int
d\mu(\eta) \log\cosh (\sqrt{\alpha \beta \bar{p}}\eta) + \\
&+& \nonumber \frac{\alpha}{2}\log (\frac{1}{1-\beta(1-\bar{q})})
+ \frac{\alpha\beta}{2}\frac{\bar{q}}{1-\beta(1-\bar{q})},
\end{eqnarray} and, overall, we can state the next
\begin{teorema}
The replica symmetric free energy of the analogical Hopfield
neural network is given by the following expression
\begin{eqnarray}\label{RS} A^{RS}(\beta,\alpha) &=& \log 2  + \int
 d\mu(\eta) \log\cosh (\sqrt{\alpha \beta \bar{p}}\eta) +
\\ &+& \nonumber \frac{\alpha}{2}\log
(\frac{1}{1-\beta(1-\bar{q})}) +
\frac{\alpha\beta}{2}\frac{\bar{q}}{1-\beta(1-\bar{q})}
-\frac{\alpha \beta}{2} \bar{p}(1 - \bar{q}) - \frac{\alpha
\beta}{2}.
\end{eqnarray}
\end{teorema}
\begin{remark} We stress that in the ergodic regime, where the
overlap self-averages to zero, the expression recover the correct
ergodic expression \cite{bg1} as well as the annealed expression
of the Sherrington-Kirkpatrick model (SK) when sending $\alpha \to
\infty$ and $\beta \to 0$ by keeping  $\alpha \beta = \beta_{SK}$.
\end{remark}

Self-consistency relations can be found by imposing equal to zero
the partial derivatives of the free energy with respect to its
order parameters, namely the system $(\partial_q A(\beta,\alpha) =
0), (\partial_p A(\beta,\alpha) = 0)$, which gives
\begin{eqnarray}\label{selfq}
\frac{\partial A}{\partial q} &=& \frac{\alpha \beta}{2} \Big( \bar{p} -
\frac{\beta \bar{q}}{(1-\beta(1-q))^2} \Big) = 0\\
\frac{\partial A}{\partial p} &=& \frac{\alpha \beta}{2} \Big(
\int  d\mu(z)\tanh^2(\sqrt{\alpha \beta \bar{p}}z) - \bar{q} \Big)
= 0 \label{selfp},
\end{eqnarray}
by which \be \bar{q} = \int  d\mu(z)
\tanh^2\Big( \frac{\sqrt{\alpha \bar{q}}\beta
z_{\mu}}{(1-\beta(1-\bar{q}))} \Big), \ee and as a consequence
$\bar{p}(\bar{q})= \beta \sigma^4 \bar{q}$.
These conditions can be seen as a minimax principle defining the replica symmetric
solution. Let us recall that in the spin glass case we have a minimum principle instead \cite{Gsum}.
\section{Fluctuations of the order parameters and critical
line}\label{tre}

We are now ready to separate different regions in the phase
diagram, where different behaviors do appear. In particular we
want to see where the annealing, characterized by $(q=0,p=0)$, is
spontaneously broken and ergodicity is lost.
\newline
To satisfy this task we  proceed as follows: at first we introduce
the streaming equation so to be able to calculate variations of
generic observable as overlap correlation functions.
\newline
Then we define the centered and rescaled overlaps and introduce
their correlation matrix. Each element of this matrix then is
evaluated at $t=0$ and then propagated thought $t=1$ via its
streaming: This procedure encodes naturally for a system of
coupled linear differential equations that, once solved, give the
expressions of the overlap fluctuations. The latter are found to
diverge on a line in the $(\alpha,\beta)$ plane, which becomes a
natural candidate for a second order phase transition (confirmed
by the regularity of the behavior before such a line is reached
from the ergodic phase).
\newline
Let us start the plan by introducing the following
\begin{proposition}
Given $O$ as a smooth function of $s$ replica overlaps
$(q_1,...,q_s)$ and $(p_1,...,p_s)$, the following streaming
equation holds:
\begin{eqnarray} \label{streaming} \frac{d}{dt}\langle O \rangle_t
&=& \beta \sqrt{\alpha} \Big( \sum_{a,b}^s \langle O \cdot
\xi_{a,b}\eta_{a,b} \rangle_t \\ \nonumber &-& s \sum_{a=1}^s
\langle O \cdot \xi_{a,s+1}\eta_{a,s+1} \rangle_t +
\frac{s(s+1)}{2}\langle O \cdot \xi_{s+1,s+2}\eta_{s+1,s+2}
\rangle_t \Big). \end{eqnarray}
\end{proposition}
We skip the proof as is long but simple and works by direct
evaluation pretty standard in the disordered system literature
(see for example \cite{Gsum,barra1,bg2}).

\bigskip

The rescaled overlap $\xi_{12}$ and $\eta_{12}$ are defined
accordingly to
\begin{eqnarray}
\xi_{12}= \sqrt{N}\Big( q_{12} - \bar{q} \Big), \\
\eta_{12}= \sqrt{K}\Big( p_{12} - \bar{p} \Big).
\end{eqnarray}

In order to control the overlap fluctuations, namely $\langle
\xi_{12}^2 \rangle_{t=1}$, $\langle \xi_{12}\eta_{12}
\rangle_{t=1}$, $\langle \eta_{12}^2 \rangle_{t=1}$,  ..., noting that
the streaming equation pastes two replicas to the ones already
involved ($s=2$ so far), we need to study nine correlation
functions. It is  then useful to introduce them and link them to
capital letters so to simplify their visualization:

\begin{eqnarray}
\langle \xi_{12}^2 \rangle_t &=& A(t), \ \ \  \langle
\xi_{12}\xi_{13}
\rangle_t = B(t), \ \ \ \langle \xi_{12}\xi_{34}\rangle_t = C(t), \\
\langle \xi_{12}\eta_{12} \rangle_t &=& D(t), \ \ \ \langle
\xi_{12}\eta_{13} \rangle_t = E(t), \ \ \ \langle
\xi_{12}\eta_{34}\rangle_t = F(t), \\
\langle \eta_{12}\eta_{12} \rangle_t &=& G(t), \ \ \ \langle
\eta_{12}\eta_{13} \rangle_t = H(t), \ \ \ \langle
\eta_{12}\eta_{34}\rangle_t = I(t).
\end{eqnarray}

Let us now sketch their streaming. Let us at first introduce the
operator ``dot'' as
$$
\dot{O}  = \frac{1}{\beta\sqrt{\alpha}}\frac{d O}{dt},
$$
which simplifies calculations and shifts the propagation of the
streaming from $t=1$ to $t=\beta\sqrt{\alpha}$: Using it we sketch
how to write the streaming of the first two correlations (as it
works in the same way for any other):
\begin{eqnarray} \nonumber
\dot{A} &=& \langle \xi_{12}^2 \xi_{12}\eta_{12} \rangle_t -4
\langle \xi_{12}^2\xi_{13}\eta_{13} \rangle_t + 3 \langle
\xi_{12}^2 \xi_{34}\eta_{34} \rangle_t, \\ \nonumber \dot{B} &=&
\langle \xi_{12}\xi_{13}\Big( \xi_{12}\eta_{12} +
\xi_{13}\eta_{13} + \xi_{23}\eta_{23} \Big)\rangle_t - \\ &-& 3
\langle \xi_{12}\eta_{13} \Big( \xi_{14}\eta_{14} +
\xi_{24}\eta_{24} + \xi_{34}\eta_{34} \Big)\rangle_t  + 6 \langle
\xi_{12}\eta_{13} \xi_{45}\eta_{45} \rangle_t. \nonumber
\end{eqnarray}
By assuming a Gaussian behavior, as in the strategy outlined in \cite{Gsum}, we can write the overall  streaming of the correlation
functions in the form of  the following differential system
\begin{eqnarray}\nonumber
\dot{A} &=& 2AD - 8BE + 6CF, \\ \nonumber \dot{B} &=& 2AE + 2BD -
4BE - 6BF - 6EC + 12CF, \\ \nonumber \dot{C} &=& 2AF + 2CD + 8BE -
16BF - 16CE + 20CF, \\ \nonumber \dot{D} &=& AG - 4BH + 3CI + D^2
-4E^2 + 3F^2, \\ \nonumber \dot{E} &=& AH+BG -2BH -3BI -3CH + 6CI
+ 2ED -2E^2 -6EF + 6F^2, \\ \nonumber\dot{F} &=& AI + CG + 4BH
-8BI -8 CH + 10 CI  + 2DF + 4E^2 -16EF + 10F^2, \\ \nonumber
\dot{G} &=& 2GD - 8HE + 6IF, \\ \nonumber \dot{H} &=& 2GE + 2HD -
4HE - 6HF -6IE + 12IF, \\ \nonumber \dot{I} &=& 2GF + 2DI + 8HE -
16 HF - 16IE + 20IF.
\end{eqnarray}
It is easy to solve this system, once the initial conditions at $t=0$ are known. Our general analysis covers also the case where external fields are involved. We do not report here the full analysis, for the sake of brevity.  

In order to proceed further, in our case of absence of external fields, we need to evaluate these correlations at
$t=0$. As at $t=0$ everything is factorized, the only needed check
is by the correlations inside each party.

Starting with the first party, we have to study $A,B,C$ at $t=0$.
As only the diagonal terms give not negligible contribution, it is
immediate to work out this first set of starting points as

\begin{eqnarray}
A(0) &=&  N^{-1}\sum_i^N (1- 2\bar{q}\langle \sigma_i^{1}\sigma_i^{2} \rangle + \bar{q}^2) = 1 - \bar{q}^2, \\
B(0) &=& N^{-1}\sum_i^N (\sigma_i^2 \sigma_i^3 - \bar{q}\sigma_i^1 \sigma_i^2 - \bar{q}\sigma_i^1 \sigma_i^3 +
\bar{q}^2)  = \bar{q} - \bar{q}^2, \\
C(0) &=&
N^{-1}\sum_{ij}^{N,N}(\sigma_i^{1}\sigma_i^{2}\sigma_i^{3}\sigma_i^{4}-
\bar{q} \sigma_i^{1}\sigma_i^{2} -\bar{q} \sigma_j^{3}\sigma_j^{4}
 + \bar{q}^2) = \nonumber \\ &=& \int d\mu(z) \tanh^4(\frac{\beta \sqrt{\alpha \bar{q}}
z}{1-\beta(1-\bar{q})}) - \bar{q}^2,
\end{eqnarray}
where we stress that even in the last equation only the diagonal
terms $i=j$ contribute. \newline For the second party we need to
evaluate $G,H,I$ at $t=0$. The only difference with the first
party is the lacking of the dichotomy of its elements such that
$z_{\mu}^2 \neq 1$ as for the $\sigma$'s.

It is immediate to check that $G(0), H(0), I(0)$ are function of
$\omega(z^2)$ and $\omega^2(z)$, which are Gaussian integrals and
can be we worked out as
\begin{eqnarray}
\omega(z) &=& \frac{\int z e^{\sqrt{\beta \bar{q}}\eta z}
e^{\frac{\beta}{2}(1-\bar{q})z^2}e^{-z^2/2}dz}{\int e^{\sqrt{\beta
\bar{q}}\eta z} e^{\frac{\beta}{2}(1-\bar{q})z^2}e^{-z^2/2}dz} =
\sqrt{\beta \bar{q}} \eta \sigma^2, \\
\omega(z^2) &=& \frac{\int z e^{\sqrt{\beta \bar{q}}\eta z}
e^{\frac{\beta}{2}(1-\bar{q})z^2}e^{-z^2/2}dz}{\int e^{\sqrt{\beta
\bar{q}}\eta z} e^{\frac{\beta}{2}(1-\bar{q})z^2}e^{-z^2/2}dz} =
\sigma^2(1+\sqrt{\beta \bar{q}}\eta\sigma)^2.
\end{eqnarray}
Remembering that $\beta \sigma^4\bar{q}=\bar{p}$ (cfr.
eq.(\ref{selfp})), we get
\begin{eqnarray}\nonumber
G(0) &=& \mathbb{E}\omega(z^2)\omega(z^2) - \bar{p}^2=
\mathbb{E}\sigma^4(1+\sqrt{\beta \bar{q}}\sigma\eta)^4 - \bar{p}^2, \\
\nonumber H(0) &=& \mathbb{E}\omega(z^2)\omega(z)^2 -\bar{p}^2=
\mathbb{E}\sigma^2(1+\sqrt{\beta \bar{q}}\eta\sigma)^2\beta\bar{q}\eta^2\sigma^4-\bar{p}^2, \\
\nonumber I(0) &=& \mathbb{E}\omega^4(z) - \bar{p}^2 =
\mathbb{E}(\beta\bar{q})^2\eta^4\sigma^8-\bar{p}^2.
\end{eqnarray}
The last step missing is averaging over the $\eta$, by exploiting $\langle
\eta^2 \rangle=1$, $\langle \eta^4 \rangle=3$.  Finally, we have obviously $D(0)=E(0)=F(0)=0$, 
because at $t=0$ the two parties are independent. 

Here, we are interested in finding where
ergodicity becomes broken (the critical line),  we start
propagating  $t \in 0 \to 1$ from the annealed region, where $\bar{q}
\equiv 0$ and $\bar{p}\equiv 0$.
\newline
It is immediate to check that, for the only terms that we need to
consider, $A,D,G$ (the other being strictly zero on the whole
$t\in [0,1]$), the starting points are $A(0)=1, D(0)=0,
G(0)=(1-\beta)^{-2}$ and their evolution is ruled by
\begin{eqnarray}
\dot{A} &=& 2AD, \\ \dot{D} &=& AG + D^2, \\ \dot{G} &=& 2GD.
\end{eqnarray}
So we need to solve the system above. The first step is noticing
that
$$
d_t \log A = \frac{\dot{A}}{A} = 2D =  \frac{\dot{G}}{G} = d_t
\log G,
$$
as $d(A/G)/dt = 0$, and $A(0)/G(0)=(1-\beta)^2$, we obtain
immediately the coupled behavior of the self-correlations: \be
A(t) = G(t)(1-\beta)^2. \ee We now reduced to consider the system
\begin{eqnarray}\label{flu1}
\dot{D} &=& (1-\beta)^2 G^2 + D^2, \\ \label{flu2}\dot{G} &=& 2GD.
\end{eqnarray}
Let us call $[D+(1-\beta)G]=Y$ such that summing (\ref{flu1}) and
(\ref{flu2}) we get the differential equation
$$
\dot{Y}(t) = Y^2(t) \Rightarrow Y(t) = \frac{Y_0}{1-t Y_0},
$$
by which, as $Y_0 = (1-\beta)^{-1}$, we get \be\label{flu_coupled}
D(t=\sqrt{\alpha}\beta) + (1-\beta) G(t=\sqrt{\alpha}\beta) =
\frac{1}{1-\beta(1+\sqrt{\alpha})}, \ee i.e. there is a regular
behavior up to $\beta_c =1/(1 + \sqrt{\alpha})$.

\bigskip

Now, starting from eq.(\ref{flu_coupled}),  we have to solve
separately for $G(t)$ and for $D(t)$.
\newline
Let us at first notice that \be \dot{G}(t) = 2G(t)\Big(Y(t)
-(1-\beta)G(t) \Big), \ee by which, dividing both the sides by
$G^2$ and considering $Z=G^{-1}$, we get \be -\dot{Z}(t)
-2Y(t)Z(t) + 2(1-\beta)=0, \ee namely an ordinary first order
differential equation for $Z(t)$.
\newline
We solve it by posing $Z(t) = W(t)\exp\Big( -2\int_0^t Y(t')dt'
\Big)$, with $Z_0=W_0$ fixing the auxiliary function $W(t)$ as
$$
\int_0^t Y(t')dt' = \log\Big( \frac{1-\beta}{1-\beta -t} \Big).
$$
We can obtain in a few algebraic steps the function $Z(t)$ and
consequently, remembering that $G(t)=Z^{-1}(t)$ we get
\be\label{flu_G} G(t)= \frac{1}{2(1-\beta)}\Big( \frac{1}{1-\beta
-t}  + \frac{1}{1-\beta +t}  \Big) = \frac{1}{(1-\beta)^2 - t^2}.
\ee Now it is possible to insert eq.(\ref{flu_G}) into
(\ref{flu_coupled}) which concludes the proof of  the following
\begin{teorema}
In the ergodic region the behavior of the overlap fluctuations is
regular and described by  the following equations:
\begin{eqnarray}
\langle \xi_{12}^2 \rangle &=& \frac{(1-\beta)^2}{(1-\beta)^2 -
\beta^2 \alpha}, \\
\langle \xi_{12}\eta_{12} \rangle &=& \frac{
\beta\sqrt{\alpha}}{(1-\beta)^2 - \beta^2 \alpha}, \\
\langle \eta_{12}^2 \rangle &=& \frac{1}{(1-\beta)^2 - \beta^2
\alpha}.
\end{eqnarray}
The ergodic region ends in the line \be \beta_c=
\frac{1}{1+\sqrt{\alpha}}, \ee which is the critical line.
\newline We stress that it turns out to be the same AGS-line of
the standard neural network counterpart.
\end{teorema}

\section{Conclusion and outlook}\label{quattro}

In this paper we achieved another step toward a general theory of
neural networks whose statistical mechanics is not based on
replica-trick.
\newline
We found the replica symmetric behavior of the analogical Hopfield
model, its self-averaging equations for the order parameters and a
complete quantitative picture of their fluctuations and
correlations. The critical line defining ergodicity breaking is
found as well, in agreement with the standard AGS counterpart.
\newline
Furthermore the method paves the way for analytical investigation
of general bipartite systems, which are assuming by themselves a
very important role in applied statistical mechanics
\cite{contucci1}.
\newline
Despite these new results, fundamental enquiries are still open:
apart the challenging thermodynamic limit, the retrieval phase
(the response to an external stimulus) has not been discussed so
far, neither the replica symmetry breaking scheme, which should be
incorporated in the theory too.
\newline
We plan to report soon on these topics.

\bigskip
\noindent
{\bf Acknowledgements}
\newline
Support from MiUR (Italian Ministry of University and Research)
and INFN (Italian Institute for Nuclear Physics) is gratefully
acknowledged.
\newline
AB work is supported by the SmartLife Project (Ministry Decree
$13/03/2007$ n.$368$) which is acknowledged.

\end{document}